\documentclass[letter]{aa}
\usepackage{graphicx}
\usepackage[varg]{txfonts}
\usepackage{natbib}
\bibpunct{(}{)}{;}{a}{}{,}
\usepackage{array}
\usepackage{xspace}
\usepackage{lscape}
\usepackage{url}

\usepackage[hyperfootnotes=false, linktocpage=true, breaklinks=true, colorlinks=true, linkcolor=blue, citecolor=blue, urlcolor=blue]{hyperref}
\usepackage[all]{hypcap}

\usepackage{mathtools}
\usepackage[figuresright]{rotating}

\newcommand{\mgxi}{\ion{Mg}{xi}\xspace}
\newcommand{\sxvi}{\ion{S}{xvi}\xspace}
\newcommand{\kxviii}{\ion{K}{xviii}\xspace}
\newcommand{\sixiv}{\ion{Si}{xiv}\xspace}
\newcommand{\arxvii}{\ion{Ar}{xvii}\xspace}
\newcommand{\arxviii}{\ion{Ar}{xviii}\xspace}
\newcommand{\clxvi}{\ion{Cl}{xvi}\xspace}
\newcommand{\clxvii}{\ion{Cl}{xvii}\xspace}
\newcommand{\caxix}{\ion{Ca}{xix}\xspace}

\newcommand{\fexvii}{\ion{Fe}{xvii}\xspace}

\newcommand{\fexxv}{\ion{Fe}{xxv}\xspace}

\begin{document}

\title{A novel scenario for the possible X-ray line feature at $\sim$3.5 keV:}
\subtitle{Charge exchange with bare sulfur ions}

\author {Liyi Gu \inst{1} 
\and 
Jelle Kaastra \inst{1,2} 
\and 
A. J. J. Raassen\inst{1,3} 
\and 
P. D. Mullen\inst{4}
\and
R. S. Cumbee\inst{4}
\and 
D. Lyons\inst{4}
\and
P. C. Stancil\inst{4}}

\institute{
SRON Netherlands Institute for Space Research, Sorbonnelaan 2, 3584 CA Utrecht, The Netherlands \\ \email{L.Gu@sron.nl}
\and 
Leiden Observatory, Leiden University, PO Box 9513, 2300 RA Leiden, The Netherlands 
\and 
Astronomical Institute ``Anton Pannekoek'', Science Park 904, 1098 XH Amsterdam, University of Amsterdam, The Netherlands 
\and
Department of Physics and Astronomy and the Center for Simulational Physics, University of Georgia, Athens, GA 30602, USA }

%==============%
\abstract
%==============%
{
Motivated by recent claims of a compelling $\sim$3.5 keV emission line from nearby galaxies and galaxy clusters, 
we investigate a novel plasma model incorporating a charge exchange component obtained from theoretical scattering
calculations. Fitting this kind of component with a 
standard thermal model yields positive residuals around 3.5 keV, produced mostly by \sxvi transitions from principal quantum numbers $n \geq 9$ to the ground.
Such high-$n$ states can only be populated by the charge exchange process. In this scenario, the observed 3.5 keV line flux in clusters 
can be naturally explained by an interaction in an effective volume of $\sim$1 kpc$^3$ between a $\sim$3 keV temperature plasma 
and cold dense clouds moving at a few hundred km s$^{-1}$. The \sxvi lines at $\sim$3.5 keV also provide a unique diagnostic 
of the charge exchange phenomenon in hot cosmic plasmas.
}


\authorrunning{L. Gu et al.}
\titlerunning{\sxvi Charge Exchange Lines at $\sim 3.5$ keV}

\maketitle

%====================%
\section{Introduction}
%====================%

Recently, a narrow X-ray band of around 3.5 keV has attracted unprecedented attention in the astrophysics and particle
physics communities. Bulbul et al. (2014; hereafter BU14) and Boyarsky et al. (2014; hereafter BO14) independently reported  
discoveries of an unidentified line feature (hereafter ULF) at $\sim$3.5 keV in the X-ray spectra of bright galaxy
clusters and the Andromeda galaxy, respectively. Because the nearby thermal emission lines (e.g., \kxviii at 3.51 keV) from hot astrophysical plasmas are
expected to be much weaker than the observed value, BU14 and BO14 suggested that the ULF could stem from the decay of $\sim 7$ keV sterile 
neutrino dark matter particles.

These claims ignited lively debates on two basic issues: (1) is the ULF real; and (2) if real, is it indeed a feature due to nonbaryonic matter.     
Several follow-up studies claimed positive detections of a ULF, although the reported line central energies are not fully reconciled with each other
(e.g., Urban et al. 2015, hereafter U15; Iakubovskyi et al. 2015). Others, e.g., Malyshev et al. (2014), Anderson et al. (2014), Tamura et al. (2015),
Carlson et al. (2015), and Sekiya et al. (2015), were unable to confirm the existence of the ULF and give only upper limits. Several detections and nondetections
were claimed with exactly the same datasets, indicating that the ULF search is subject to great complexity and it is still premature to answer
issue (1). This leads us to  study issue (2) instead to scrutinize the interpretations of a possible 3.5 keV ULF from an astrophysical context.

We propose a novel scenario that the 3.5 keV ULF can be explained to some extent by charge exchange (hereafter CX) processes 
between hot cosmic plasma and cold neutral gas. Astronomical CX X-rays were first discovered in near-Earth comets (Lisse et al. 1996)
and now the related detections have been expanded to various types of objects, from neighboring planets to distant galaxy clusters (see the 
review by Dennerl et al. 2010). A qualitative discussion on the possible CX lines near 3.5 keV was presented in BU14. Compared to 
their model, our new plasma code, as described in detail in Gu et al. (2015), is expected to produce more 
realistic line emissivities. Our code incorporates the velocity-dependent cross-section data with resolved final states based on the most
up-to-date theoretical calculations.

This letter is organized as follows. In \S2, we
introduce our CX model and describe the characteristic feature at $\sim$3.5 keV. In \S3, we compare our model with the actual ULF observations and summarize the new novel scenario in \S4.

%====================%
\section{Charge exchange recombination of bare sulfur ions}
%====================%

%====================%
\subsection{High-$n$ capture by charge exchange}
%====================%
%============================
%  FIG: four combined figure
%
\begin{figure*}[!]
\centering
\resizebox{0.6\hsize}{!}{\includegraphics[angle=0]{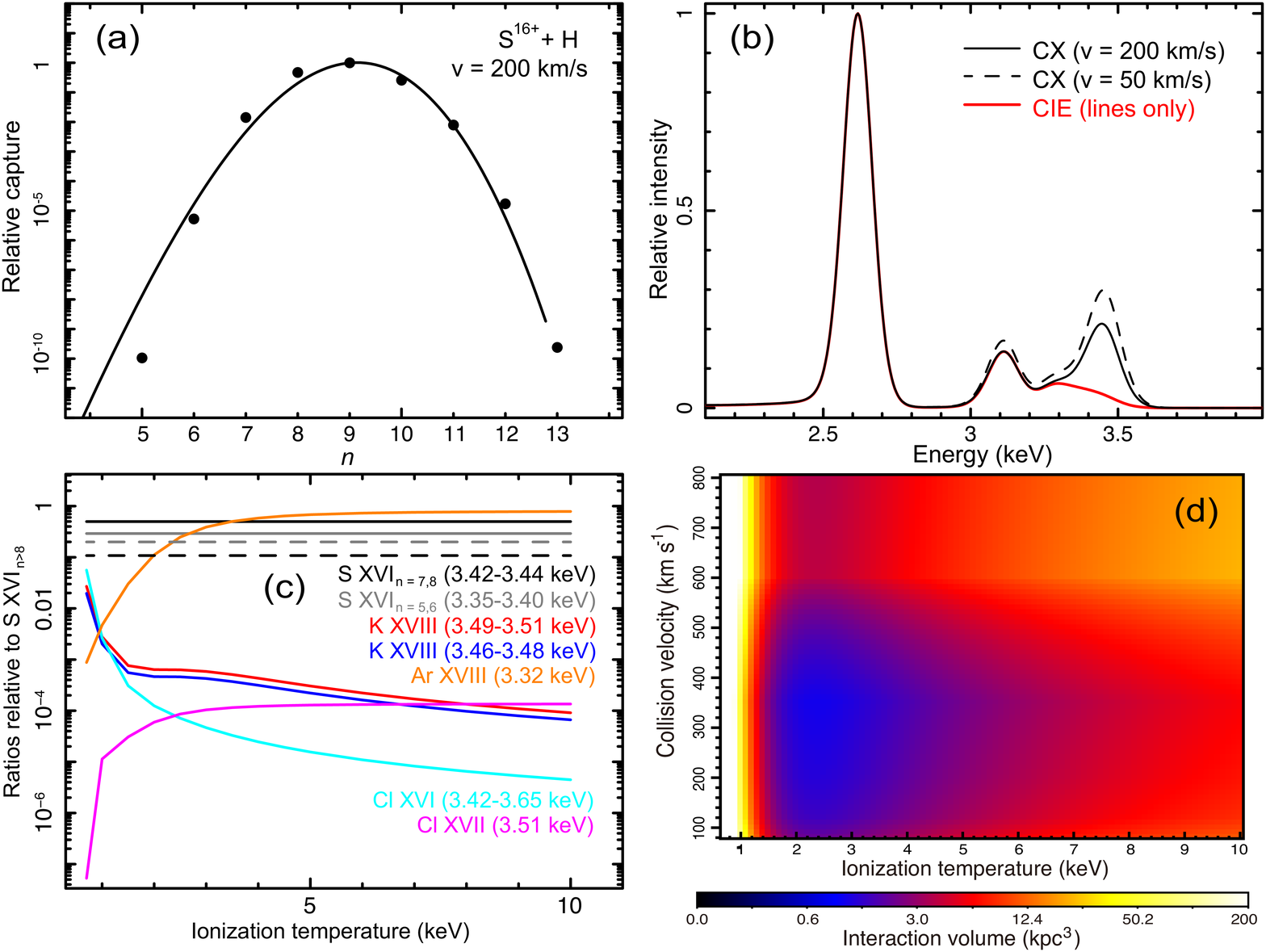}}
\caption{(a) Fractions of electron captured into $n$ shells at $v=200$ km $\rm s^{-1}$ for S$^{16+}$ + H. The solid line is the scaling relation obtained in 
Gu et al. (2015), and the data points come from actual calculation by Cumbee et al. (2015). (b) \sxvi line emission, normalized to the 
Ly$\alpha$ transition for CX (black) and CIE (red) models. The CX spectra with $v = 200$ km s$^{-1}$ and 50 km s$^{-1}$ are shown in 
solid and dashed lines, respectively. The ionization temperatures are set to 3 keV in all cases, and the spectra are folded with the 
{\it XMM-Newton} PN response. (c) All the CX lines in the $3.3-3.7$ keV band, normalized to the \sxvi transitions from $n>8$ to the ground 
at $3.45-3.47$ keV, and plotted as a function of ionization temperature.        The solid lines are obtained by setting $v = 200$ km s$^{-1}$, 
while for \sxvi transitions, we also plot, in dashed lines, the results with $v = 50$ km s$^{-1}$. (d) A 2-D color map of interaction volume $V$, 
on the plane of ionization temperature and collision velocity, to reproduce the reported ULF flux (\S 3.2).
}
\label{four_fig}
\end{figure*}
%============================

Charge exchange occurs when multicharged ions collide with neutrals, leaving 
product ions with one additional electron in an excited state. Radiative relaxation 
of the product ions create unique line emission spectra (e.g., different from collisional excitation
or radiative recombination), providing a novel
probe to the interactions between hot and cold astrophysical plasmas. As described
in Gu et al. (2015), we have developed a new plasma code, as an independent model in the SPEX package (Kaastra et al. 1996),\footnote{https://www.sron.nl/spex} to calculate X-ray emission due to charge 
exchange. The flux is given by

\begin{equation}
F = \frac{1}{4 \pi D_{\rm l}^{2}} \int n_{\rm I} n_{\rm N} v \sigma_{\rm I, N}(v,n,l,S) dV,
\end{equation}      
\noindent where $D_{\rm l}$ is the luminosity distance of the object, $n_{\rm I}$ and $n_{\rm N}$
are the densities of ionized and neutral media, respectively, $v$ is the collisional
velocity, $\sigma_{\rm I, N}$ is the charge exchange cross section, and $V$ is the interaction volume. The cross
section is usually in excess of $10^{-15}$ cm$^{2}$, much larger than those of other recombination processes. For each
reaction, $\sigma_{\rm I, N}$ highly depends on $v$, and exhibits strong selective behavior on product ion subshells with quantum numbers
$n$, $l$, and $S$ (see the review by Janev \& Winter 1985). This is because the CX probability usually reaches its maximum when the internal energy 
of the initial and final electron quantum states are in near resonance. As shown in Fig. 1a, for a collision with 
$v=200$ km s$^{-1}$ between 
fully stripped sulfur and atomic hydrogen,
${\bf S}^{\bf 16+} + {\bf H} \rightarrow {\bf S}^{\bf 15+}(n,l,S) + {\bf H}^{\bf +}$,
the capture probability has its maximum at subshell $n=9$, and the adjacent subshells $n=$7, 8, 10, and 11 are also 
populated to $\sim 1-50$\% of the peak value. The CX rates of these subshells are substantially larger than the excitation rates. At a lower velocity,
e.g., $v = 50$ km s$^{-1}$, the dominant level becomes $n=10$ instead and the $n-$distribution is more peaked than the high
velocity case. The most recent cross-section data by Cumbee et al. (2015) were fully implemented in our model.

The highly excited states subsequently decay to the ground state. 
As described in Gu et al. (2015), the CX spectrum was then calculated by modeling the radiative cascade network of the recombining ions. 
For all atomic levels up to $n=16$ and $l=15$, we have obtained the energies and transition probabilities from multiple databases
(e.g., National Institute of Standards and Technology) and codes (e.g., Flexible Atomic Code; Gu 2008). For the \sxvi ion, the
dominant transitions from $n=9$ and 10 are 9$p$ $\rightarrow$ 1$s$ at 3.45 keV, and 10$p$ $\rightarrow$ 1$s$ at 3.46 keV, respectively.
These central energies, as well as the transition rates, are consistent with those used in other plasma codes (e.g., AtomDB; Foster et al. 2012).

%====================%
\subsection{Lines of \sxvi  near 3.5 keV}
%====================%

Similar to the emission by collisional excitation with free electrons, CX can also create strong
Ly$\alpha$ lines, as the $n=2$ shell would be effectively populated by radiative cascade. When the upper level $n$ 
increases, however, the CX and CIE (collisional ionization equilibrium) spectra gradually separate from each other.
As shown in Fig. 1b, the \sxvi CX emission apparently exceeds that of the
CIE at around 3.45 keV, although the same ionization temperature (3 keV) is 
assumed for both models. After convolving with a typical CCD response, the excess becomes a broadened feature between 3.4 keV and 3.6 keV.
This kind of feature is more apparent at a lower collision velocities regime, where the CX population becomes more focused on a single
outer shell (\S2.1). The comparison between CX and CIE indicates that the CX emission of \sxvi might create a positive residual near
3.5 keV when it is fitted by the CIE model.

To compare  the intensities of all CX lines in the $3.3-3.7$ keV band quantitatively, we calculated 
the emissivities of \sxvi transitions from $n>4$ to the ground, as well as other related ions, 
including \arxviii lines at 3.32 keV, \kxviii at $3.46-3.51$ keV, \clxvi at $3.42-3.65$ keV,
and \clxvii at 3.51 keV. A solar abundance ratio (Lodders et al. 2009) was assumed for the four elements and the CX collision velocity was set to 200 km s$^{-1}$. 
Then we normalized all the transitions 
to the strongest one, i.e., \sxvi transitions from $n>8$ to the ground at $3.45-3.47$ keV. The relative emissivities are plotted in Fig. 1c
as a function of ionization temperature. For \sxvi, we also calculated line ratios at $v = 50$ km s$^{-1}$. The \sxvi transitions at
$3.45-3.47$ keV, as illustrated in Fig. 1c, are always the most prominent CX feature around 3.5 keV; other ions have lower 
cosmic abundances, and/or weaker transitions in this band. 
%============================
%  FIG: resdiual figures
%
\begin{figure*}[!]
\centering
\resizebox{1.\hsize}{!}{\hspace{-1cm}\includegraphics[angle=0]{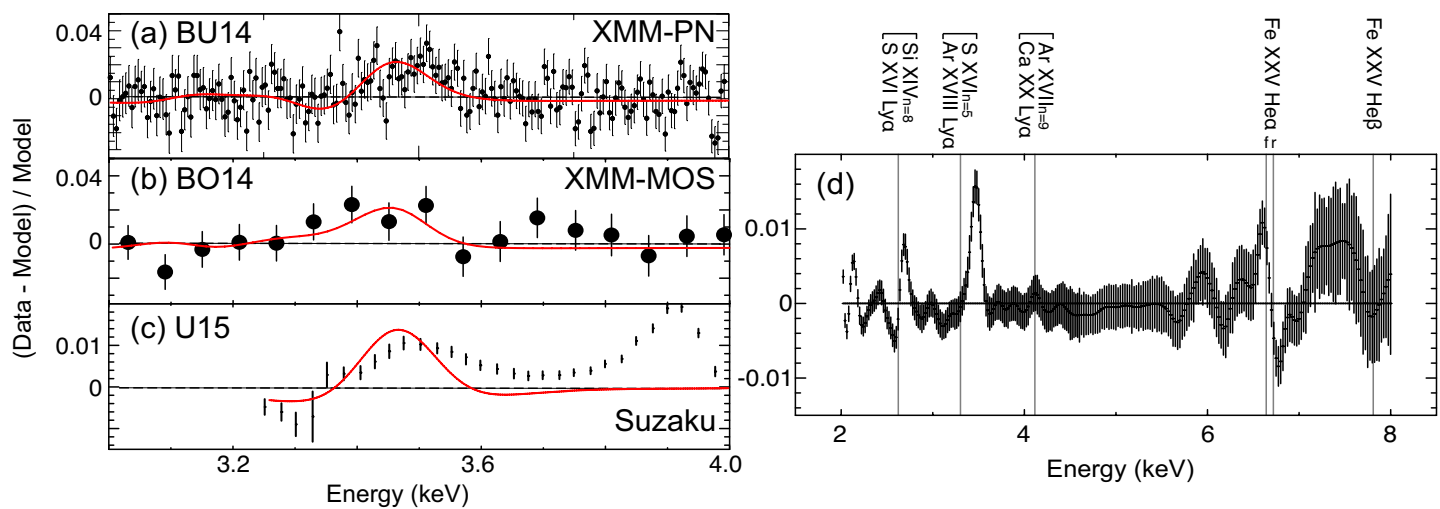}}
\caption{(a, b, and c) The spectral fitting residuals reported in BU14 (top right panel of their Fig. 5), BO14 (right panel of their Fig. 6), and U15 (left panel of their Fig. 2), 
  together with the expected residuals, shown as red curves, from our scenario described in \S3.2. Emission measure of the CX model in (c) is half of those in (a) and (b). As described in \S3.2,
    the data-to-model deviation at $\sim 3.9$ keV in (c) might be caused by other plasma emission lines.  (d) Fitting residual from 
a 1 Ms simulation of the {\it XMM-Newton} PN observation of the Perseus cluster. Details are described in \S3.3. Vertical lines indicate the typical blended CIE and CX lines and the \fexxv transitions in the 6.6-7.9 keV.
}
\label{res_fig}
\end{figure*}
%============================

%====================%
\section{Application to the reported line at $\sim$3.5 keV}
%====================%
Based on our own calculation, we propose a new emission line near 3.5 keV from charge exchange between bare sulfur ions and neutral hydrogen atoms. 
The CX line center is close to that of the ULF reported by BU14, BO14, U15, and Iakubovskyi et al. (2015). 
Next, we examine whether or not the CX line is responsible for the ULF. This immediately poses three questions to 
be addressed. (i) Assuming the ULF is indeed created by CX, what is the origin of the CX emission? (ii) Are the properties (e.g., flux and central energy) 
of the predicted CX line consistent with the 
ULF observation? And (iii) why, given that CX lines are expected for many ions, is only the \sxvi feature detected? Below we examine these issues one by one.
%====================%
\subsection{Charge exchange in galaxy clusters and galaxies}
%====================%

We consider that the CX emission originates from galaxy clusters by the hot tenuous intracluster medium (ICM) penetrating and interacting with cold dense clouds.
The cold clouds might dwell around central galaxies (e.g., Conselice et al. 2001), and/or may be entrained by member galaxies 
during their infall (e.g., Gu et al. 2013a). The former produce CX emission focused on the cluster core, while the latter would give CX mostly in the outer regions, as 
a great portion of the cold content in member galaxies would be stripped out owing to, e.g., ICM ram pressure, before falling into the cluster center 
(e.g., Gu et al. 2013b). Recently, Walker et al. (2015) reported that CX might contribute a large fraction of X-ray flux in the 
$0.5-1.0$ keV band emitted from H$\alpha$ filaments around central galaxies.
Since the ICM is highly ionized, it is natural to expect \sxvi lines from the same filaments. This kind of cluster CX scenario can further explain the nondetection
of the ULF in the center of the Virgo cluster (e.g., BU14 and U15), where little cold gas was found (a few $10^{6}$ $M_{\odot}$, Salom$\rm \acute{e}$ \& Combes 2008).
In contrast, the Perseus cluster, often regarded as a template example for positive ULF detection, harbors a $\sim$1.7$\times 10^{10}$ $M_{\odot}$ cold cloud in the center (Edge 2001).

This scenario can be also applied to the central regions of spiral galaxies. As reported in Koyama et al (1989, 1996), a large amount of hot plasma
was found in our Galactic center, emitting K-shell lines from various highly-ionized ions, including \sxvi. A similar concentration of hot plasma, though 
relatively dimmer, was discovered in  M31 (e.g., Takahashi et al. 2004). Such regions are also rich in cold dense clouds (e.g., Morris \& Serabyn 1996).

%====================%
\subsection{Comparison with the ULF observations}
%====================%

Now we address question (ii) based on the cluster CX scenario. To explain the observed ULF flux ($4 \times 10^{8}$ photons s$^{-1}$ cm$^{-2}$, BU14), we calculated the required emission measure $n_{\rm I} n_{\rm N} V$ for various temperatures and velocities based on Eq.~1. By adopting the observed mean ICM density of 
$3 \times 10^{-2}$ cm$^{-3}$ (Perseus cluster center; Churazov et al. 2003) and a typical cold cloud density of 10~cm$^{-3}$ (e.g., Heiner et al. 2008), the required 
interaction volume was obtained and plotted in Fig. 1d as a function of temperature and velocity. 
This result demonstrates that a $V \sim 1$ kpc$^{3}$ interaction can reproduce the observed flux, if the ionization temperature is mild ($\sim 2-4$ keV) and $v$ is less than $\sim 500$ km s$^{-1}$.  
This kind of plasma temperature is well in line with the observed values in the Perseus cluster (e.g., Churazov et al. 2003). The expected collision velocity agrees with a combination of the cold
cloud velocity, thermal motion of sulfur ions, and ICM bulk/turbulence motion, which are observed to be $0-400$ km s$^{-1}$ (Salom$\rm \acute{e}$ et al. 2006), $\sim$100$-$200 km~s$^{-1}$ (for the ICM temperature of $2-5$ keV),
and $\sim$100 km~s$^{-1}$ (Zhuravleva et al. 2014) in the Perseus case, respectively. The derived interaction volume can be interpreted as a portion of the H$\alpha$ filament, which has a 
scale of several 10 kpc (Salom$\rm \acute{e}$ et al. 2006). As shown in Lallement et al. (2004), the thickness of CX layer is of the same order as the mean free path of hydrogen atom against collisional ionization. Assuming typical ICM parameters, we then obtain an interaction depth of $\sim 0.1$ kpc and surface of $\sim 10$ kpc$^2$. Hence, the simple cluster CX scenario can naturally explain the observed ULF flux.

The central energy of the ULF varies up to $\sim$100 eV among different detections (e.g., $3.51-3.57$ keV in BU14; $3.46-3.53$ keV in BO14). To
compare directly with observations, we simulated the fitting residual of the 3.5 keV structure and plot it in Figs. 2a, 2b, and 2c, together with the observed residuals from BU14, BO14, and U15, respectively. We use the stacked EPIC-PN data from BU14 instead of their MOS data, since the latter has a known caveat on an \arxvii line at 3.62 keV (see \S6 of BU14). For BO14 and U15, we focus on the Perseus data. 
The simulated spectra consist of multiple line-free CIE components (four for BU14, and two for B014 and U15), whose parameters were set to the observed values of BU14 (in their Table 2), and a CX component with solar abundance. The temperature and velocity of the CX model were set to 2.5 keV and 200 km s$^{-1}$, respectively. After folding the 
spectra with the detector responses, we fit them, in the $3-4$ keV band, with the same model as used in BU14, namely four line-free CIE components plus five Gaussian lines for the known thermal transitions. As shown in Fig. 2a, 2b, and 2c, the cluster CX scenario can approximately reproduce the observed ULF in the $3.4-3.55$ keV band; the remaining residuals become $\leq 1$\% of the model. The excess at $\sim 3.9$ keV in the U15 data is probably due to the model uncertainty of the \caxix He$\alpha$ transitions (\S4.3 of U15).

%====================%
\subsection{Why was the ULF only detected at $\sim$3.5 keV?}
%====================%

We then consider question (iii), i.e., why the \sxvi transitions at $\sim$3.5 keV are unique in terms of the ULF detection.
To examine possible CX features at other energies, we simulated {\it XMM-Newton} spectra with both CIE and CX components and 
fitted them with CIE alone. The temperature and flux of the central region of Perseus 
cluster reported in U15 were utilized for the CIE component, and CX was characterized by a typical ionization temperature
of 2.5 keV, $v = 200$ km s$^{-1}$ and an effective interaction volume of 1 kpc$^{3}$. Solar abundances were used for both
components. The simulated spectra were fitted with a CIE model consisting of four temperature components (as in BU14), and the best-fit residuals in the $2-8$ keV band are plotted 
in Fig. 2d. The $<$2 keV band is ignored in all the ULF searches, as it is too crowded with lines. The residuals 
appear as weak line structures with amplitudes $<$2\% of the model. The \sxvi lines at $\sim$3.5 keV are clearly seen, while the CX lines of other ions 
become less apparent, since they
often overlap somewhat in energy with certain strong thermal lines in a typical CCD spectrum. For example, the strong CX line of \sixiv 
from $n=8$ to the ground has an energy of 2.63 keV, which cannot be distinguished from the thermal \sxvi Ly$\alpha$ transition at 2.62 keV.
Similar conditions would apply to the \fexvii CX lines in the soft band, as the characteristic transitions from $n \geq 8$ to the ground overlap with the thermal \mgxi
lines at $\sim 1.35$ keV. On the other hand, the \fexxv features shown in the $6.6-7.9$ keV band (see, for example, Mullen et al. 2015) are relatively isolated, but become more vulnerable to the uncertainties of the instrumental background. 
In contrast, the 3.5 keV band is free of strong thermal lines and instrumental issues, making the \sxvi transitions a unique feature for probing CX in hot astrophysical plasmas.

%====================%
\subsection{Solar wind charge exchange}
%====================%

CX emission is often seen when solar wind ions collide with interstellar and geocoronal neutrals. If the ULF is created alternatively by this kind of 
local solar wind charge exchange (hereafter SWCX), it would only vary with the solar wind conditions. Unfortunately, such SWCX contamination is often difficult to remove 
by a standard X-ray light curve filtering (e.g., Snowden et al. 2008). By analyzing the archival data of the Solar Wind Electron Proton Alpha
Monitor onboard the Advanced Composition Explorer, we determined the mean proton flux for the period of each X-ray exposure 
for Perseus, M31, Virgo center, and two samples of dwarf galaxies and normal galaxies used in BU14, BO14, U15, Malyshev et al. (2014), and Anderson et al. (2015).
As shown in Fig. 3, all reports claiming positive ULF detection are indeed based on the data, which were partly taken in the relatively high proton flux periods,
while the two nondetection cases, i.e., Virgo center and stacked dwarf galaxies, were observed at relatively quiescent periods with the proton flux
$\leq 1.5 \times 10^{8}$ cm$^{-2}$ s$^{-1}$. The nondetection case of the stacked normal galaxy sample (Anderson et al. 2015), which covers a large range in the proton flux, is an
apparent outlier to the SWCX scenario.

Another challenge to this scenario is the required high charge state. The average freeze-in ionization temperature of solar wind ions
are commonly below $2 \times 10^{6}$ K (Feldman et al. 2005), suggesting a negligible ion fraction of S$^{16+}$. Although a high charge
state SWCX with $\geq 1$ keV temperature may exist (e.g., Carter et al. 2010), it appears to be associated with an extreme coronal mass ejection
event, which is rather rare.

\begin{figure}
\begin{center}
\includegraphics[angle=-0,scale=.3]{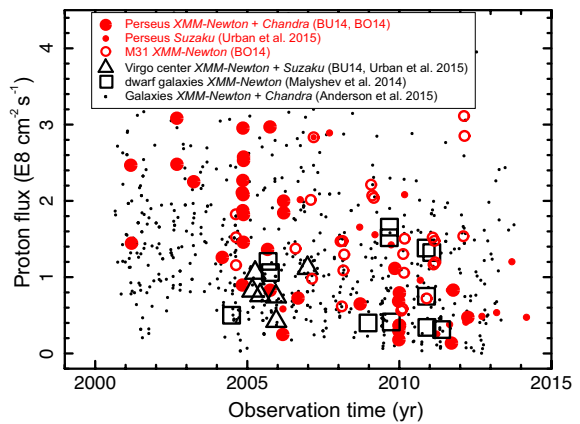}
\caption{Solar wind proton flux, taken from the ACE SWEPAM, at periods of X-ray observations of the Perseus cluster, Virgo cluster center, M31, and two samples of dwarf galaxies and normal galaxies. The
  ULF detection and nondetection cases were plotted in red and black, respectively. }
\end{center}
\end{figure}

%====================%
\section{Summary}
%====================%

By calculating the line emission due to charge exchange with cold H gas, we propose a novel scenario that the reported X-ray line at $\sim$3.5 keV in clusters and galaxies
can be explained by a set of \sxvi transitions from $n \geq 9$ to the ground. Such high-$n$ transitions can be produced exclusively
by charge transfer between bare sulfur ions and neutral particles. Both ingredients are naturally present in galaxy clusters and central regions of spiral galaxies.
The \sxvi transitions 3.5 keV are a unique feature for CX detection with current instruments; other CX lines are expected to 
be observed with the future {\it Astro-H} mission.

\section*{Acknowledgments}
SRON is supported financially by NWO, the Netherlands Organization for Scientific Research. Work at UGA was partially supported
by NASA grants NNX09AC46G and NNX13AF31G.

\end{document}